\documentstyle[12pt,epsfig]{article}
\def\lbldef#1#2{\expandafter\gdef\csname #1\endcsname {#2}}

\def\href#1#2{#2}  

\begin{document}
\baselineskip=15.5pt
\pagestyle{plain}
\setcounter{page}{1}

\begin{titlepage}

\begin{flushright}
CERN-TH/2000-126\\
hep-th/0004144
\end{flushright}
\vspace{10 mm}

\begin{center}
{\Large Extra Force in Brane Worlds}

\vspace{5mm}

\end{center}

\vspace{5 mm}

\begin{center}
{\large Donam Youm\footnote{E-mail: Donam.Youm@cern.ch}}

\vspace{3mm}

Theory Division, CERN, CH-1211, Geneva 23, Switzerland

\end{center}

\vspace{1cm}

\begin{center}
{\large Abstract}
\end{center}

\noindent

By carefully analyzing the geodesic motion of a test particle in the 
bulk of brane worlds, we identify an extra force which is recognized 
in spacetime of one lower dimensions as a non-gravitational 
force acting on the particle.  Such extra force acts on the particle 
in such a way that the conventional particle mechanics in one lower 
dimensions is violated, thereby hinting at the higher-dimensional 
origin of embedded spacetime in the brane world scenario.  We obtain 
the explicit equations describing the motion of the bulk test particle 
as observed in one lower dimensions for general gravitating 
configurations in brane worlds and identify the extra non-gravitational 
force acting on the particle measured in one lower dimensions.

\vspace{1cm}
\begin{flushleft}
CERN-TH/2000-126\\
April, 2000
\end{flushleft}
\end{titlepage}
\newpage

\section{Introduction}

There has been renewed interest in compactification through the 
non-factorizable Kaluza-Klein (KK) metric Ansatz with the warp factor, 
after Randall and Sundrum (RS) showed \cite{rs1,rs2,rs3} that 
such unconventional compactification of the extra spatial dimensions 
provides with a simple solution to the hierarchy problem of particle 
physics.  A novel and surprising feature of the RS scenario is that even 
if the extra spatial dimension is infinite in size, the Newton's $1/r^2$ 
law of four-dimensional gravity is recovered with negligible correction 
from the massive KK modes of graviton.  The previous works on gravitational 
aspects of the RS scenario have attempted to reproduce physics of 
four-dimensional gravity up to small corrections beyond the current 
experimental precision in an effort to provide with an evidence that the 
RS scenario may be a true description of our nature.  

Especially, in Refs. \cite{hwa,youm1,cceh} the geodesic equation for 
a massless test particle in the bulk of a gravitating configuration 
in brane world is put into the form of the geodesic motion of a massive 
test particle in the corresponding gravitating configuration in one lower 
dimensions by parameterizing the geodesic path with an affine 
parameter associate with the canonical metric in one lower dimensions.  
Such result appears to indicate that the bulk geodesic motion is 
observed by a four-dimensional observer to reproduce physics of our 
four-dimensional spacetime.  However, as we will see through the careful 
analysis of the geodesic motion of a test particle in the bulk spacetime 
of brane worlds, in general laws of physics governing the motion of a particle 
in four-dimensional spacetime is observed to be violated.  Namely, we find 
that the equation describing the trajectory of a particle as observed in one 
lower dimensions of the brane world scenario has an extra force term which 
is parallel to the velocity of the particle.  According to the so-far known 
four-dimensional physics, only the component of the extra non-gravitational 
four-force $F^{\mu}$ which is perpendicular to the particle's four-velocity 
${{dx^{\mu}}\over{d\tau}}$ can influence the particle's motion.  Due to such 
unusual property which cannot be explained by the physics of four-dimensional 
spacetime, such extra force was dubbed in the previous literature 
\cite{gp,cp,wes1,wes2,wes3,wes4,wes5,wes6} as the fifth force.  
[Such extra abnormal force is observed also in Refs. \cite{sch,kov,gk} 
by analyzing the geodesic motion of a test particle in the five-dimensional 
Kaluza-Klein theory.]  The so-called fifth force generically exists in the KK 
theories (with constant moduli scalar fields of the extra space) when the 
spacetime metric depends on the extra spatial coordinates and the velocity 
of the particle has nonzero components in both the extra spatial direction 
and the direction of our three dimensional space.  Therefore, it is 
inevitable that if our four-dimensional world is described by the RS 
scenario then a four-dimensional observer following a test particle 
along its bulk geodesic path should observe the violation of four-dimensional 
law of particle mechanics, since the RS scenario allows dependence of the 
spacetime metric on the extra spatial coordinate and physical process in the 
RS scenario is generally higher-dimensional in nature.  

The paper is organized as follows.  In section 2, we survey relevant aspects 
of domain wall solutions in and geodesic motion of a test particle in the 
bulk of the RS scenario.  In this section, we also discuss well-known facts 
of particle mechanics in curved spacetime for the purpose of understanding 
the physical implication of the extra force observed in one lower dimensions. 
Although some aspects of the fifth force were already studied in the previous 
literature, we feel that its relation to the four-dimensional particle 
mechanics has not been clearly presented.  We hope that the present paper will 
clarify some of confusing issues in the fifth force.  We study the bulk 
geodesic motion of a test particle moving in general gravitating 
configurations in brane worlds as observed in one lower dimensions for the 
case corresponding to the KK zero mode bulk graviton in section 3 and for 
the case of general bulk graviton including the massive KK modes in section 
4.  In these sections, we also identify the extra force on the particle which 
is observed in one lower dimensions as non-gravitational.  Conclusions are 
given in section 5.

\section{Preliminaries and General Setup}

In this section, we prepare for the main topic of this paper by surveying 
aspects of domain wall solutions and dynamics of a particle in brane 
worlds and in general relativity.

Generally, the spacetime metric
\footnote{In this paper, we use the mostly positive convention $(-+\cdots+)$ 
for the metric signature.} 
for a $D$-dimensional domain wall can be put into the following form:
\begin{equation}
G_{MN}dx^Mdx^N={\cal W}(y)\eta_{\mu\nu}dx^{\mu}dx^{\nu}+dy^2,
\label{warpmet}
\end{equation}
where $M,N=0,1,...,D-1$, $\mu,\nu=0,1,...,D-2$ and ${\cal W}(y)$ is the 
warp factor.  In particular, for the domain wall solution to the 
field equations of the following bulk action:
\begin{equation}
S={1\over{2\kappa^2_D}}\int d^Dx\sqrt{-G}\left[{\cal R}_G-{4\over{D-2}}
\partial_M\phi\partial^M\phi+e^{-2a\phi}\Lambda\right],
\label{dilact}
\end{equation}
the warp factor is given by \cite{rs1,youm2,kks,youm3}
\begin{equation}
{\cal W}(y)=\left(1-{{(D-2)a^2}\over 2}\sqrt{{{D-2}\over{4(D-1)-a^2(D-2)^2}}
\Lambda}|y|\right)^{8\over{(D-2)^2a^2}},
\label{warp1}
\end{equation}
for $a\neq 0$, and 
\begin{equation}
{\cal W}(y)=\exp\left(-2\sqrt{\Lambda\over{(D-1)(D-2)}}|y|\right),
\label{warp2}
\end{equation}
for $a=0$.  Here, we have imposed the invariance under the ${\bf Z}_2$ 
transformation $y\to-y$ and chosen the warp to decrease so that the 
bulk graviton can be localized.

It is the purpose of this paper to study the dynamics of a test particle 
in the bulk spacetime with the following metric:
\begin{equation}
G_{MN}dx^Mdx^N={\cal W}g_{\mu\nu}dx^{\mu}dx^{\nu}+dy^2.
\label{bulkmet}
\end{equation}
This metric generically describes any gravitational configuration in brane 
worlds.  The geodesic motion of a test particle, i.e., the motion of a 
particle which is acted on by the gravitational force only, is described by 
the following geodesic equations:
\begin{equation}
{{d^2x^R}\over{d\lambda^2}}+\hat{\Gamma}^R_{MN}{{dx^M}\over{d\lambda}}
{{dx^N}\over{d\lambda}}=0,
\label{geodeqs}
\end{equation}
where $\hat{\Gamma}^R_{MN}$ is the Christoffel symbol (of the second 
kind) for the metric $G_{MN}$ and $\lambda$ is an affine parameter for 
the geodesic path $x^M(\lambda)$.  In addition, the metric compatibility 
along the geodesic path requires that
\begin{equation}
-\epsilon_D=G_{MN}{{dx^M}\over{d\lambda}}{{dx^N}\over{d\lambda}}=
{\cal W}g_{\mu\nu}{{dx^{\mu}}\over{d\lambda}}{{dx^{\nu}}\over{d\lambda}}
+\left({{dy}\over{d\lambda}}\right)^2,
\label{metcomp}
\end{equation}
where $\epsilon_D=1,0$ respectively for a massive test particle (i.e., a 
timelike geodesic) and a massless test particle (i.e., a null geodesic).
For a timelike geodesic, $\epsilon_D$ actually can take any positive 
values, but one can always apply the affine transformation 
$\lambda\to a\lambda+b$ ($a,b\in{\bf R}$), which leaves the geodesic 
equations (\ref{geodeqs}) invariant, to bring $\epsilon_D=1$.  In this 
paper, we shall not consider the spacelike geodesics, i.e. the $\epsilon_D=
-1$ case.  

In this paper, we shall re-express the geodesic equations (\ref{geodeqs}) 
for the bulk geodesic motion of a test particle in terms of quantities 
of the hypersurface spacetime of one lower dimensions, for the purpose of  
learning how the bulk geodesic motion of the test particle is observed in 
one lower dimensions.  In the previous related works \cite{hwa,youm1,cceh}, 
the bulk geodesic motion is reparameterized by an affine parameter 
$\tilde{\lambda}$ associated with the four-dimensional metric given by 
$g_{\mu\nu}$, namely the one satisfying the following:
\begin{equation}
-\epsilon_{D-1}=g_{\mu\nu}{{dx^{\mu}}\over{d\tilde{\lambda}}}{{dx^{\nu}}
\over{d\tilde{\lambda}}},
\label{naffin1}
\end{equation}
where $\epsilon_{D-1}=1,0$ respectively for a timelike and a lightlike motion 
observed in one lower dimensions.  On the other hand, as the bulk test 
particle follows its geodesic path, it generally passes through one 
hypersurface to another.  We note that the induced metric on the hypersurface 
at a specific value of $y$ is given by $\tilde{g}_{\mu\nu}\equiv{\cal W}(y)
g_{\mu\nu}$.  Therefore, it appears that the natural choice for the affine 
parameter $\tilde{\lambda}$ (or the standard clock on the hypersurface that 
follows the particle in its motion) for the motion observed on the 
(comoving) hypersurface is the one satisfying the following
\begin{equation}
-\epsilon_{D-1}={\cal W}(y)g_{\mu\nu}{{dx^{\mu}}\over{d\tilde{\lambda}}}
{{dx^{\nu}}\over{d\tilde{\lambda}}}.
\label{naffin2}
\end{equation}
However, one might argue that $\tilde{\lambda}$ satisfying Eq. 
(\ref{naffin1}) actually corresponds to the affine parameter for the motion 
as observed on the hypersurface $y=y_0={\rm constant}$, say the TeV brane of 
our world, at {\it fixed} distance from the Planck brane at $y=0$.  Namely, 
the affine parameter at $y=y_0$ is defined by
\begin{equation}
-\epsilon_{D-1}={\cal W}(y_0)g_{\mu\nu}(x^{\rho},y_0){{dx^{\mu}}\over
{d\tilde{\lambda}}}{{dx^{\nu}}\over{d\tilde{\lambda}}},
\label{naffin3}
\end{equation}
and by applying the affine transformation $\tilde{\lambda}\to {\cal W}
(y_0)^{-1/2}\tilde{\lambda}$, one can bring this to the form 
(\ref{naffin1}) in the case when $g_{\mu\nu}$ is independent of the 
extra spatial coordinate $y$.  But we, as beings adapted to sense only the 
four- or lower-dimensional phenomena, are incapable of looking into the 
extra spatial direction to observe objects in different four-dimensional 
hypersurface.  Any object on different four-dimensional hypersurface cannot 
be observed; it simply exists in different four-dimensional universe which 
cannot be observed.  So, the geodesic equations conveniently put into a 
four-dimensional form in term of a new affine parameter in Refs. 
\cite{hwa,youm1,cceh} actually do not describe the trajectory of a test 
particle observed in one lower dimensions, as one might naively assume 
although Refs. \cite{hwa,youm1,cceh} do not explicitly state that the 
bulk geodesic motion can be observed by an observer on the TeV brane.  
However, in the following we shall also consider the case associated 
with a choice of the affine parameter defined by Eq. (\ref{naffin1}), 
as well as the one defined by Eq. (\ref{naffin2}), just for the 
purpose of studying the general bulk geodesic motion also within the 
framework of the previous works.

In obtaining the equations for the particle motion observed in one lower 
dimensions, we assume that the affine parameter $\tilde{\lambda}$ for the 
spacetime in one lower dimensions is a smooth function of the affine parameter 
$\lambda$ for the bulk geodesic motion: $\tilde{\lambda}=f(\lambda)$.  
First, for $\tilde{\lambda}$ defined in Eq. (\ref{naffin1}), which we refer 
to as `case 1', the relation (\ref{metcomp}) for the bulk geodesic motion 
is rewritten in terms of the new parameter $\tilde{\lambda}$ as
\begin{equation}
g_{\mu\nu}{{dx^{\mu}}\over{d\tilde{\lambda}}}{{dx^{\nu}}\over
{d\tilde{\lambda}}}=-{\cal W}^{-1}\left[\epsilon_D\left({{d\lambda}\over
{d\tilde{\lambda}}}\right)^2+\left({{dy}\over{d\tilde{\lambda}}}\right)^2
\right].
\label{georeldtodm1}
\end{equation}
Second, for $\tilde{\lambda}$ defined in Eq. (\ref{naffin2}), which we refer 
to as `case 2', Eq. (\ref{metcomp}) is rewritten in terms of 
$\tilde{\lambda}$ as
\begin{equation}
{\cal W}g_{\mu\nu}{{dx^{\mu}}\over{d\tilde{\lambda}}}{{dx^{\nu}}\over
{d\tilde{\lambda}}}=-\left[\epsilon_D\left({{d\lambda}\over{d
\tilde{\lambda}}}\right)^2+\left({{dy}\over{d\tilde{\lambda}}}\right)^2
\right].
\label{cstncrel}
\end{equation}
The parameter $\tilde{\lambda}$ is an affine parameter for the motion observed 
in one lower dimensions, if Eqs. (\ref{georeldtodm1}) and (\ref{cstncrel}) 
respectively take the forms (\ref{naffin1}) and (\ref{naffin2}), i.e., 
the RHS's are either 0 or $-1$.  

We now discuss the condition for $\tilde{\lambda}$ being an affine parameter.  
First, for a massless test particle ($\epsilon_D=0$), the RHS's of Eqs. 
(\ref{georeldtodm1}) and (\ref{cstncrel}) can be 0 or $-1$, depending on 
the motion of a test particle along the $y$-direction.  When the test particle 
is confined to move along the longitudinal directions of the domain wall
\footnote{From Eq. (\ref{geodeq2}) with $\epsilon_D=0$, one can see that 
such bulk geodesic motion is possible for a massless test particle.}
(i.e., ${{dy}\over{d\lambda}}=0$ and therefore ${{dy}\over{d
\tilde{\lambda}}}={{d\lambda}\over{d\tilde{\lambda}}}{{dy}\over{d
\lambda}}=0$), the RHS's of Eqs. (\ref{georeldtodm1}) and (\ref{cstncrel}) 
are zero, corresponding to the lightlike motion as observed in one lower 
dimensions.  In this case, one can choose $\tilde{\lambda}=\lambda$ as an 
affine parameter for the motion observed in one lower dimensions, as can be 
seen from Eqs. (\ref{georeldtodm1}) and (\ref{cstncrel}).  When the 
$y$-component of the velocity of the massless test particle is nonzero 
(i.e., ${{dy}\over{d\lambda}}\neq 0$), its motion is observed in one lower 
dimensions as timelike if the following is satisfied:
\begin{equation}
\left({{dy}\over{d\tilde{\lambda}}}\right)^2={\cal W},
\label{mslsstime}
\end{equation}
for the case 1, and
\begin{equation}
\left({{dy}\over{d\tilde{\lambda}}}\right)^2=1,
\label{nulltotime}
\end{equation}
for the case 2, for which cases the RHS's of Eqs. (\ref{georeldtodm1}) and 
(\ref{cstncrel}) are $-1$.  Second, for a massive test particle 
($\epsilon_D=1$), the parameter $\tilde{\lambda}$ is an affine parameter for 
the timelike motion as observed in one lower dimensions, if $\tilde{\lambda}$ 
is related to the bulk affine parameter $\lambda$ as
\begin{equation}
\left({{d\lambda}\over{d\tilde{\lambda}}}\right)^2=
{\cal W}-\left({{dy}\over{d\tilde{\lambda}}}\right)^2,
\label{affpararel}
\end{equation}
for the case 1, and
\begin{equation}
\left({{d\lambda}\over{d\tilde{\lambda}}}\right)^2=1-\left({{dy}\over
{d\tilde{\lambda}}}\right)^2,
\label{affnpararel2}
\end{equation}
for the case 2.

In the following sections, we will see that from the perspective of an 
observer in one lower dimensions a bulk test particle appears to be under 
the influence of an abnormal non-gravitational force.  So, it would be useful 
to discuss some aspects of dynamics of particles in general relativity to 
better understand physical implication of such extra force.  Although 
we assume the spacetime to be four-dimensional in the discussion in 
the following paragraphs, our discussion holds for arbitrary spacetime 
dimensions without modification of equations.  

In terms of the relativistic four-vector notation, the following classical 
Newton's second law of mechanics plus the work-energy relation (the law of 
conservation of energy):
\begin{equation}
{\bf F}={{d{\bf p}}\over{dt}},\ \ \ \ \ \ \ \ 
{{dT}\over{dt}}={\bf F}\cdot{\bf v},
\label{newteconsv}
\end{equation}
where ${\bf F}$ is the force on the particle, ${\bf p}=m_0{\bf v}=m_0
{{d{\bf x}}\over{dt}}$ is the momentum of the particle with the 
inertial mass $m_0$ and $T$ is the kinetic energy of the particle, is 
written as 
\begin{equation}
{{dp_{\mu}}\over{d\tau}}=F_{\mu}, 
\label{momfrcrel}
\end{equation}
where $p^{\mu}=m_0{{dx^{\mu}}\over{d\tau}}$ is the contravariant components 
of the four-momentum of a particle with the rest mass $m_0$, $\tau$ is the 
proper time defined from $d\tau^2=-\eta_{\mu\nu}dx^{\mu}dx^{\nu}$,  
$F_{\mu}=(-{\bf F}\cdot{{d{\bf x}}\over{d\tau}},{\bf F}{{dt}\over{d\tau}})$, 
and $(x^{\mu})=(t,{\bf x})$.  In curved spacetime with metric $g_{\mu\nu}$, 
Eq. (\ref{momfrcrel}) is modified to:
\begin{equation}
{{Dp^{\mu}}\over{d\tau}}\equiv {{dp^{\mu}}\over{d\tau}}+
\Gamma^{\mu}_{\rho\sigma}{{dx^{\rho}}\over{d\tau}}p^{\sigma}=F^{\mu},  
\label{crvmomfrcrel}
\end{equation}
in the contravariant notation, or
\begin{equation}
{{Dp_{\mu}}\over{d\tau}}\equiv {{dp_{\mu}}\over{d\tau}}-{1\over 2}
{{\partial g_{\rho\sigma}}\over{\partial x^{\mu}}}{{dx^{\rho}}\over{d\tau}}
p^{\sigma}=F_{\mu},  
\label{crvmomfrcrel2}
\end{equation}
in the covariant notation, where $\Gamma^{\mu}_{\rho\sigma}$ is the 
Christoffel symbol of the second kind for the metric $g_{\mu\nu}$.  Here, 
$F_{\mu}$ and $p^{\mu}$ are defined in the same way as in the flat spacetime 
case except that the proper time $\tau$ is now defined from $d\tau^2=
-g_{\mu\nu}dx^{\mu}dx^{\nu}$.  

First, we show that a purely mechanical non-gravitational four-force 
$F^{\mu}$ acts on a particle perpendicularly to its four-velocity ${{dx^{\mu}}
\over{d\tau}}$.  By taking the covariant derivative of $g_{\mu\nu}{{dx^{\mu}}
\over{d\tau}}{{dx^{\nu}}\over{d\tau}}=-1$ along the particle trajectory 
$x^{\mu}(\tau)$, one can see that $g_{\mu\nu}{D\over{d\tau}}\left(
{{dx^{\mu}}\over{d\tau}}\right){{dx^{\nu}}\over{d\tau}}=0$.  This implies that 
$g_{\mu\nu}F^{\mu}{{dx^{\nu}}\over{d\tau}}=g_{\mu\nu}{{Dp^{\mu}}\over{d\tau}}
{{dx^{\nu}}\over{d\tau}}=m_0g_{\mu\nu}{D\over{d\tau}}\left({{dx^{\mu}}
\over{d\tau}}\right){{dx^{\nu}}\over{d\tau}}=0$.  

Next, we consider the possibility that the four-force $F^{\mu}$ has nonzero 
component parallel to the four-velocity ${{dx^{\mu}}\over{d\tau}}$. 
In deriving the relation $g_{\mu\nu}F^{\mu}{{dx^{\nu}}\over{d\tau}}=0$ 
in the previous paragraph, we assumed that the proper mass $m_0$ of the 
particle is constant in time.  Had we considered the possibility that $m_0$ 
changes with time, we would instead have obtained
\begin{equation}
g_{\mu\nu}F^{\mu}{{dx^{\nu}}\over{d\tau}}=g_{\mu\nu}{{Dp^{\mu}}\over{d\tau}}
{{dx^{\nu}}\over{d\tau}}=m_0g_{\mu\nu}{D\over{d\tau}}\left({{dx^{\mu}}\over
{d\tau}}\right){{dx^{\nu}}\over{d\tau}}+{{dm_0}\over{d\tau}}g_{\mu\nu}
{{dx^{\mu}}\over{d\tau}}{{dx^{\nu}}\over{d\tau}}=-{{dm_0}\over{d\tau}}.
\label{relwthcht}
\end{equation}
Therefore, the existence of parallel component of the force $F^{\mu}$ 
implies non-conservation of the proper mass $m_0$ of a particle.  Generally, 
the change in the proper mass of a particle occurs when there exist some 
non-mechanical external forces which cause such change.  To take into 
account of the additional non-mechanical forces, one has to modify $F_0$ 
in the following way:
\begin{equation}
F_0=-{\bf F}\cdot{{d{\bf x}}\over{d\tau}}-Q{{dt}\over{d\tau}},
\label{zerfcomp}
\end{equation}
where the first term on the RHS is the mechanical work done by the mechanical 
force ${\bf F}$ per unit time and the second term is the heat or 
non-mechanical energy developed per unit time.  The extra term is added to 
take into account of the contribution from the non-mechanical energy so that 
the energy can be conserved in the system under consideration.  
Meanwhile, the remaining components of $F_{\mu}$ take the same form as the 
above, i.e., (the curved space analog of) the Newton's second law of 
mechanics ${\bf F}={{d{\bf p}}\over{dt}}$ continues to hold.  So, from Eqs. 
(\ref{relwthcht}) and (\ref{zerfcomp}), we obtain
\begin{equation}
{{dm_0}\over{d\tau}}=Q\left({{dt}\over{d\tau}}\right)^2,
\label{heattomass}
\end{equation} 
namely, the proper mass is converted into the non-mechanical energy and 
vice versa. 

Finally, we discuss the motion of a particle under the influence of an 
extra non-gravitational force $F^{\mu}$ in curved spacetime.  First, when 
$F^{\mu}$ acts on the particle perpendicularly to its four-velocity 
${{dx^{\mu}}\over{d\tau}}$, i.e., the proper mass $m_0$ of the particle 
is constant in time, from Eqs. (\ref{crvmomfrcrel}) and (\ref{crvmomfrcrel2}) 
we obtain the following equations for the particle trajectory $x^{\mu}(\tau)$:
\begin{eqnarray}
{{d^2x^{\mu}}\over{d\tau^2}}+\Gamma^{\mu}_{\rho\sigma}{{dx^{\rho}}\over
{d\tau}}{{dx^{\sigma}}\over{d\tau}}={{F^{\mu}}\over{m_0}},
\cr
{{d^2x_{\mu}}\over{d\tau^2}}-{1\over 2}{{\partial g_{\rho\sigma}}\over
{\partial x^{\mu}}}{{dx^{\rho}}\over{d\tau}}{{dx^{\sigma}}\over{d\tau}}=
{{F_{\mu}}\over{m_0}}.
\label{traj1}
\end{eqnarray}
As expected, when the particle is free, i.e., when only gravitational force 
is acted on the particle ($F^{\mu}=0$), the particle will execute geodesic 
motion.  Second, when the force $F^{\mu}$ has nonzero parallel component, 
i.e., when $m_0$ is not conserved, the equation for the particle trajectory 
$x^{\mu}(\tau)$ takes the following form:
\begin{equation}
{{d^2x^{\mu}}\over{d\tau^2}}+\Gamma^{\mu}_{\rho\sigma}{{dx^{\rho}}\over
{d\tau}}{{dx^{\sigma}}\over{d\tau}}=
{{F^{\mu}}\over{m_0}}+g_{\rho\sigma}{{F^{\rho}}\over{m_0}}{{dx^{\sigma}}
\over{d\tau}}{{dx^{\mu}}\over{d\tau}},
\label{unfmtnextf2}
\end{equation} 
which can be obtained from Eqs. (\ref{crvmomfrcrel}) and (\ref{relwthcht}).  
We note that the RHS of Eq. (\ref{unfmtnextf2}) is perpendicular to the 
four-velocity ${{dx^{\mu}}\over{d\tau}}$, implying that only the perpendicular 
component of $F^{\mu}$ influences the motion of the particle.  In other words, 
if $F^{\mu}$ is parallel to ${{dx^{\mu}}\over{d\tau}}$, then the particle will 
just follow the geodesic path (since the RHS of Eq. (\ref{unfmtnextf2}) 
vanishes for this case) while its mass changing with time according to the 
relation (\ref{heattomass}).  This result also shows that according to 
the particle mechanics of four-dimensional general relativity the extra 
non-gravitational force term in the equation for the particle trajectory 
cannot have non-zero component parallel to the four-velocity of the particle.

\section{Dynamics in the Kaluza-Klein Zero Mode Spacetime}

In this section, we consider the case when $g_{\mu\nu}$ in the bulk metric 
(\ref{bulkmet}) does not depend on the extra spatial coordinate $y$.  
In this case, the bulk test particle is regarded as being under the influence 
of the Kaluza-Klein zero mode of graviton, only.

The geodesic equations (\ref{geodeqs}), with the bulk metric given by 
Eq. (\ref{bulkmet}) with $g_{\mu\nu}=g_{\mu\nu}(x^{\rho})$, take the 
following forms:
\begin{equation}
{{d^2x^{\rho}}\over{d\lambda^2}}+\Gamma^{\rho}_{\mu\nu}{{dx^{\mu}}\over
{d\lambda}}{{dx^{\nu}}\over{d\lambda}}+{{\cal W}^{\prime}\over{\cal W}}
{{dx^{\rho}}\over{d\lambda}}{{dy}\over{d\lambda}}=0,
\label{warpgeodeqs1}
\end{equation}
\begin{equation}
{{d^2y}\over{d\lambda^2}}-{1\over 2}{\cal W}^{\prime}g_{\mu\nu}{{dx^{\mu}}
\over{d\lambda}}{{dx^{\nu}}\over{d\lambda}}=0,
\label{warpgeodeqs2}
\end{equation}
where $\Gamma^{\rho}_{\mu\nu}$ is the Christoffel symbol for the 
metric $g_{\mu\nu}$ and ${\cal W}^{\prime}\equiv d{\cal W}/dy$.  
By using the relation (\ref{metcomp}), one can put the $y$-component 
geodesic equation (\ref{warpgeodeqs2}) into the following form:
\begin{equation}
{{d^2y}\over{d\lambda^2}}+{1\over 2}{{{\cal W}^{\prime}}\over{\cal W}}
\left[\epsilon_D+\left({{dy}\over{d\lambda}}\right)^2\right]=0.
\label{geodeq2}
\end{equation}

\subsection{Case 1}

In this subsection, we study the motion of the bulk test particle as 
observed in one lower dimensions with the spacetime metric $g_{\mu\nu}
=g_{\mu\nu}(x^{\rho})$.  

First, we consider the geodesic motion of a massless particle in the bulk 
spacetime, i.e. the $\epsilon_D=0$ case.  When ${{dy}\over{d\lambda}}=0$, 
by choosing $\tilde{\lambda}=\lambda$ as an affine parameter, one can 
put the $x^{\rho}$-component bulk geodesic equation (\ref{warpgeodeqs1}) 
into the following form:
\begin{equation}
{{d^2x^{\rho}}\over{d\tilde{\lambda}^2}}+\Gamma^{\rho}_{\mu\nu}
{{dx^{\mu}}\over{d\tilde{\lambda}}}{{dx^{\nu}}\over{d\tilde{\lambda}}}=0.
\label{nultonul}
\end{equation}
So, the test particle's motion is observed in one lower dimensions with the 
metric $g_{\mu\nu}$ as the lightlike geodesic motion.  Next, when the 
$y$-component of the velocity of the test particle is nonzero (i.e., 
${{dy}\over{d\lambda}}\neq 0$), from  Eqs. (\ref{mslsstime}) and 
(\ref{geodeq2}) with $\epsilon_D=0$, one can see that the parameters 
$\lambda$ and $\tilde{\lambda}$ are related as
\begin{equation}
\left({{d\tilde{\lambda}}\over{d\lambda}}\right)^{-1}
{d\over{d\tilde{\lambda}}}\left({{d\tilde{\lambda}}\over{d\lambda}}\right)=
-{\cal W}^{-{1\over 2}}{\cal W}^{\prime},
\label{relofparas}
\end{equation}
which can be solved by
\begin{equation}
{{d\tilde{\lambda}}\over{d\lambda}}={\cal W}^{-1}.
\label{pararelmslsstime}
\end{equation}
By using Eqs. (\ref{mslsstime}) and (\ref{relofparas}), one can 
express the $x^{\rho}$-component bulk geodesic equation (\ref{warpgeodeqs1}) 
in terms of the $(D-1)$-dimensional affine parameter $\tilde{\lambda}$.  
The resulting equation also takes the form (\ref{nultonul}).  To summarize, 
a massless test particle moving in the bulk spacetime with the metric 
(\ref{bulkmet}) is observed in one lower dimensions with the metric 
$g_{\mu\nu}(x^{\rho})$ to be ($i$) a {\it free massless} particle if the 
motion of the test particle is confined along the longitudinal directions 
of the domain wall and ($ii$) a {\it free massive} particle if the 
$y$-component of its velocity is nonzero.  This is just a generalization 
of the result in Ref. \cite{hwa} to the case of an arbitrary warp factor 
${\cal W}$ and an arbitrary gravitating configuration with the metric 
$g_{\mu\nu}(x^{\rho})$ within the brane world.  (See also Ref. \cite{cceh} 
for the generalization to the case of multi-codimensional brane world.)  

Second, we consider the geodesic motion of a massive bulk test particle, 
i.e., the $\epsilon_D=1$ case.  By using the relation (\ref{affpararel}), 
one can re-express the $y$-component geodesic equation (\ref{geodeq2}) (with 
$\epsilon_D=1$) in terms of the new parameter $\tilde{\lambda}$ as follows:
\begin{equation}
{{d^2y}\over{d\tilde{\lambda}^2}}-{{{\cal W}^{\prime}}\over{\cal W}}
\left({{dy}\over{d\tilde{\lambda}}}\right)^2+{1\over 2}{\cal W}^{\prime}=0.
\label{geodeqnewparay}
\end{equation}
Making use of Eqs. (\ref{affpararel}) and (\ref{geodeqnewparay}), one can 
again put the $x^{\rho}$-component bulk geodesic equation (\ref{warpgeodeqs1}) 
into the form (\ref{nultonul}).  So, we see that a massive test particle 
moving in the bulk spacetime with the metric (\ref{bulkmet}) is observed in 
one lower dimensions with the metric $g_{\mu\nu}(x^{\rho})$ to be a {\it free} 
massive particle.  This result extends the previous studies 
\cite{hwa,youm1,cceh} on the geodesic motion in brane worlds to include the 
case of a massive test particle.  [Note, even if some aspects of the geodesic 
motion of a massive test particle in brane worlds have been previously 
studied, it has never been shown that the geodesic motion of a {\it massive} 
test particle in the bulk spacetime is observed in one lower dimensions as 
the motion of a {\it free massive} particle.]

Just by looking at Eq. (\ref{nultonul}), it appears from the lower-dimensional 
perspective that the particle is under the influence of the gravitational 
field $g_{\mu\nu}(x^{\rho})$, only.  However, this is not the case, as we 
explain in the following.  A test particle of mass $m_0$ in $D$-dimensional  
bulk spacetime with the metric (\ref{bulkmet}) appears in $(D-1)$-dimensional 
embedded spacetime with the metric $g_{\mu\nu}$ to have mass given by 
\cite{sch,sch2}:
\begin{equation}
\tilde{m}_0=m_0{{d\tilde{\lambda}}\over{d\lambda}}=
m_0\left[{\cal W}-\left({{dy}\over{d\tilde{\lambda}}}\right)^2
\right]^{-{1\over 2}},
\label{ffmss}
\end{equation}
where we used Eq. (\ref{affpararel}).  So, as the test particle executes 
its geodesic motion with the nonzero $y$-component velocity ${{dy}\over
{d\tilde{\lambda}}}$, its mass appears to change with the following 
rate from the $(D-1)$-dimensional perspective:
\begin{equation}
{{d\tilde{m}_0}\over{d\tilde{\lambda}}}=-\tilde{m}_0{{{\cal W}^{\prime}}
\over{\cal W}}{{dy}\over{d\tilde{\lambda}}}.
\label{mchrt}
\end{equation}
From Eqs. (\ref{nultonul}) and (\ref{mchrt}), we obtain the following 
equation describing the conservation of energy and the Newton's second law 
of mechanics in $(D-1)$-dimensions:
\begin{equation}
{{Dp^{\mu}}\over{d\tilde{\lambda}}}\equiv {{dp^{\mu}}\over{d\tilde{\lambda}}}
+\Gamma^{\mu}_{\rho\sigma}{{dx^{\rho}}\over{d\tilde{\lambda}}}p^{\sigma}=
-\tilde{m}_0{{{\cal W}^{\prime}}\over{\cal W}}{{dy}\over
{d\tilde{\lambda}}}{{dx^{\mu}}\over{d\tilde{\lambda}}},
\label{frcl}
\end{equation}
where $p^{\mu}=\tilde{m}_0{{dx^{\mu}}\over{d\tilde{\lambda}}}$ is the 
$(D-1)$-momentum of the particle.  This implies that from the 
$(D-1)$-dimensional perspective the particle is under the influence of 
the extra non-gravitational velocity dependent force $F^{\mu}=
-\tilde{m}_0{{{\cal W}^{\prime}}\over{\cal W}}{{dy}\over
{d\tilde{\lambda}}}{{dx^{\mu}}\over{d\tilde{\lambda}}}$.  This extra 
force does not influence the motion of the particle, because it acts 
parallelly to the particle's $(D-1)$-velocity ${{dx^{\mu}}\over
{d\tilde{\lambda}}}$ (Cf. Eq. (\ref{unfmtnextf2})), but is responsible 
for the change of the inertial mass $\tilde{m}_0$ of the particle (from 
the $(D-1)$-dimensional perspective).  The intuitive reason is that the 
effect of the force ${\bf F}$ on the particle velocity ${\bf v}={{d{\bf x}}
\over{dt}}$ is canceled by the effect of the inertial mass change with 
time on ${\bf v}$.  Although the extra force is gravitational in nature from 
the $D$-dimensional perspective, a $(D-1)$-dimensional observer will sense 
that some non-mechanical non-gravitational force causes the inertial mass 
$\tilde{m}_0$ of the particle to be converted into the heat energy and 
vice versa.  Such heat energy generated per unit time is
\begin{equation}
Q{{dt}\over{d\tilde{\lambda}}}=-\tilde{m}_0{{{\cal W}^{\prime}}\over
{\cal W}}{{dy}\over{dt}},
\label{heatgen}
\end{equation}
according to Eq. (\ref{heattomass}).

\subsection{Case 2}

In this subsection, we study the motion of a bulk test particle as 
observed on the (comoving) hypersurface with the spacetime metric 
$\tilde{g}_{\mu\nu}={\cal W}(y)g_{\mu\nu}(x^{\rho})$.

First, we consider the bulk geodesic motion of a massless particle 
($\epsilon_D=0$).  When the motion of the test particle is confined 
along the longitudinal directions of the domain wall (i.e., 
${{dy}\over{d\lambda}}=0$), one can choose $\tilde{\lambda}=\lambda$ as an 
affine parameter on the hypersurface $y={\rm constant}$ to express the 
$x^{\rho}$-component bulk geodesic equations (\ref{warpgeodeqs1}) in 
the following form:
\begin{equation}
{{d^2x^{\rho}}\over{d\tilde{\lambda}^2}}+\tilde{\Gamma}^{\rho}_{\mu\nu}
{{dx^{\mu}}\over{d\tilde{\lambda}}}{{dx^{\nu}}\over{d\tilde{\lambda}}}=0,
\label{nultonul2}
\end{equation}
where $\tilde{\Gamma}^{\rho}_{\mu\nu}$ is the Christoffel symbol for 
the metric $\tilde{g}_{\mu\nu}={\cal W}g_{\mu\nu}$ and we used the 
fact that $\tilde{\Gamma}^{\rho}_{\mu\nu}=\Gamma^{\rho}_{\mu\nu}$.  
So, its motion observed on the hypersurface $y={\rm constant}$ is that of 
a massless free particle.  Nontrivial result arises when the massless test 
particle has nonzero $y$-component for its velocity.  From Eqs. 
(\ref{nulltotime}) and (\ref{geodeq2}) with $\epsilon_D=0$, one can 
see that the parameters $\lambda$ and $\tilde{\lambda}$ are related as
\begin{equation}
\left({{d\tilde{\lambda}}\over{d\lambda}}\right)^{-1}
{d\over{d\tilde{\lambda}}}\left({{d\tilde{\lambda}}\over{d\lambda}}\right)=
-{1\over 2}{{{\cal W}^{\prime}}\over{\cal W}},
\label{parasrel}
\end{equation}
which can be solved by
\begin{equation}
{{d\tilde{\lambda}}\over{d\lambda}}={\cal W}^{-{1\over 2}}.
\label{paramrel2}
\end{equation}
So, the bulk geodesic equation (\ref{warpgeodeqs1}) for the 
$x^{\rho}$-component motion is rewritten in terms of the new parameter 
$\tilde{\lambda}$ as
\begin{equation}
{{d^2x^{\rho}}\over{d\tilde{\lambda}^2}}+\tilde{\Gamma}^{\rho}_{\mu\nu}
{{dx^{\mu}}\over{d\tilde{\lambda}}}{{dx^{\nu}}\over{d\tilde{\lambda}}}=
-{1\over 2}{{{\cal W}^{\prime}}\over{\cal W}}
{{dx^{\rho}}\over{d\tilde{\lambda}}}.
\label{xrhoeqnull}
\end{equation}
The bulk geodesic motion of a free massless particle with nonzero 
$y$-component for its velocity is therefore observed on the hypersurface $y=
y(\tilde{\lambda})$ with the metric $\tilde{g}_{\mu\nu}$ as the motion of a 
massive particle which is under the influence of the extra 
non-gravitational force as well as the gravitational field 
$\tilde{g}_{\mu\nu}$.

Second, we consider the massive bulk test particle ($\epsilon_D=1$).  
By using Eq. (\ref{affnpararel2}), one can re-express the $y$-component 
geodesic equation (\ref{geodeq2}) (with $\epsilon_D=1$) in terms of the 
new parameter $\tilde{\lambda}$ as
\begin{equation}
{{d^2y}\over{d\tilde{\lambda}^2}}+{1\over 2}{{{\cal W}^{\prime}}\over
{\cal W}}\left[1-\left({{dy}\over{d\tilde{\lambda}}}\right)^2\right]=0.
\label{ydirceq}
\end{equation}
The $x^{\rho}$-component geodesic equations (\ref{warpgeodeqs1}) can be put 
into the following simplified form in terms of the new parameter 
$\tilde{\lambda}$  by applying Eqs. (\ref{affnpararel2}) and (\ref{ydirceq}):
\begin{equation}
{{d^2x^{\rho}}\over{d\tilde{\lambda}^2}}+\tilde{\Gamma}^{\rho}_{\mu\nu}
{{dx^{\mu}}\over{d\tilde{\lambda}}}{{dx^{\nu}}\over{d\tilde{\lambda}}}=
-{1\over 2}{{{\cal W}^{\prime}}\over{\cal W}}{{dy}\over
{d\tilde{\lambda}}}{{dx^{\rho}}\over{d\tilde{\lambda}}}.
\label{xdireceqs}
\end{equation}
So, the geodesic motion of a free massive particle in the bulk spacetime is 
observed on the hypersurface $y=y(\tilde{\lambda})$ with the metric 
$\tilde{g}_{\mu\nu}$ as the motion of a massive particle under the additional 
influence of an extra non-gravitational force.  For both the massive and 
massless test particle cases, the extra force term exists in the equations 
(\ref{xrhoeqnull}) and (\ref{xdireceqs}) for the particle motion, if 
the $x^{\rho}$-component of its velocity is nonzero, and acts on the 
particle parallelly to its four velocity ${{dx^{\rho}}\over
{d\tilde{\lambda}}}$.  

We have seen in the previous section that the Newton's second law of 
mechanics and the conservation of energy (and their curved space 
generalization) in four-dimensional spacetime imply that only orthogonal 
(to the four-velocity ${{dx^{\mu}}\over{d\tau}}$ of the particle) component 
of non-gravitational force $F^{\mu}$ influences the motion of the particle.  
However, as can be seen from the equations (\ref{xrhoeqnull}) and 
(\ref{xdireceqs}) for the trajectory $x^{\rho}(\tilde{\lambda})$ of the 
particle  (as observed on the comoving hypersurface $y=y(\tilde{\lambda})$), 
the particle's motion is additionally influenced by the extra force term 
(the RHS's of Eqs. (\ref{xrhoeqnull}) and (\ref{xdireceqs})) which is 
parallel to its velocity ${{dx^{\rho}}\over{d\tilde{\lambda}}}$.  Existence 
of such abnormal force term implies violation of four-dimensional physics 
and therefore can be an implication of existence of extra spatial 
dimensions (since such force term cannot be explained by the known 
four-dimensional physics).  Or maybe it is due to the wrong choice of 
frame, since we have seen in the previous subsection that in the metric 
frame of the case 1 such abnormal force term does not exist.  The author 
does not have yet clear understanding of which metric frame is the correct 
choice for the spacetime in one lower dimensions.  However, as we will see 
in the following section, when $g_{\mu\nu}$ depends on the extra spatial 
coordinate $y$, the $x^{\rho}$-component equation for the particle 
trajectory (expressed in terms of $\tilde{\lambda}$) has the extra abnormal 
force term for both case 1 and case 2.  So, it seems inevitable that the 
abnormal force term in general exists for natural choices of metric 
frame, i.e., those associated with $g_{\mu\nu}$ and $\tilde{g}_{\mu\nu}=
{\cal W}g_{\mu\nu}$.  Due to extraordinary property of such abnormal force 
term, the previous literature \cite{gp,cp,wes1,wes2,wes3,wes4,wes5,wes6} 
dubbed the extra force as the fifth force.  Actually, it should not be 
regarded as the violation of the four-dimensional physics, since 
such contradiction arises because we attempt to interpret the 
phenomenon which is higher-dimensional in nature from the perspective 
of lower dimensional physics.  We, as beings incapable of sensing 
higher-dimensional spacetime, is apt to regard the higher-dimensional 
physical process as violation of four-dimensional physics.  Anyhow, in 
the following we reconstruct the equation for the energy conservation and 
the Newton's second law of mechanics in the spacetime of one lower 
dimensions from the equations for the particle trajectory to see any 
physical implication of the extra force in one lower dimensions.

On the hypersurface $y=y(\tilde{\lambda})$ with the metric 
$\tilde{g}_{\mu\nu}={\cal W}g_{\mu\nu}$, a bulk test particle with mass 
$m_0$ appears to have mass given by
\begin{equation}
\tilde{m}_0=m_0\left[1-\left({{dy}\over{d\tilde{\lambda}}}\right)^2
\right]^{-{1\over 2}},
\label{ffmass2}
\end{equation}
where we used Eq. (\ref{affnpararel2}).  By applying Eq. (\ref{ydirceq}), 
we obtain the following mass change rate with $\tilde{\lambda}$ as 
observed on the (comoving) hypersurface:
\begin{equation}
{{d\tilde{m}_0}\over{d\tilde{\lambda}}}=-{{\tilde{m}_0}\over{2}}
{{{\cal W}^{\prime}}\over{\cal W}}{{dy}\over{d\tilde{\lambda}}}.
\label{msschrt}
\end{equation}
From Eqs. (\ref{xdireceqs}) and (\ref{msschrt}), we obtain the following 
equation describing the conservation of energy and the Newton's second 
law on the hypersurface:
\begin{equation}
{{Dp^{\mu}}\over{d\tilde{\lambda}}}\equiv {{dp^{\mu}}\over{d\tilde{\lambda}}}
+\tilde{\Gamma}^{\mu}_{\rho\sigma}{{dx^{\rho}}\over{d\tilde{\lambda}}}
p^{\sigma}=-\tilde{m}_0{{{\cal W}^{\prime}}\over{\cal W}}{{dy}\over
{d\tilde{\lambda}}}{{dx^{\mu}}\over{d\tilde{\lambda}}},
\label{phyleq}
\end{equation} 
which has the same form as the equation (\ref{frcl}) for the case 1.  
So, in both cases the particle appears to be under the influence of the 
extra force of the same form $F^{\mu}=-\tilde{m}_0{{{\cal W}^{\prime}}\over
{\cal W}}{{dy}\over{d\tilde{\lambda}}}{{dx^{\mu}}\over{d\tilde{\lambda}}}$.  
However, the motions of the particle are different for the two cases, because 
the mass $\tilde{m}_0$ changes with different rates due to different choices 
of the $(D-1)$-dimensional metric.  For the case 2, the particle mechanics 
of one lower dimensions appears to be violated, whereas for the case 1 it 
is not.

\section{Dynamics in General Spacetime}

In this section, we consider the case when $g_{\mu\nu}$ in the bulk metric 
(\ref{bulkmet}) depends on the extra spatial coordinate $y$. 
In this case, a bulk test particle is regarded as being under the influence 
of both the zero and the massive KK modes of graviton.  We will see 
that the massive KK modes of graviton induce the perpendicular 
component of the extra force $F^{\mu}$ for both cases 1 and 2, and 
the parallel component of the force term even for the case 1.  

The bulk geodesic equations (\ref{geodeqs}), with the bulk metric given by 
Eq. (\ref{bulkmet}) with $g_{\mu\nu}=g_{\mu\nu}(x^{\rho},y)$, take the 
following forms:
\begin{equation}
{{d^2x^{\rho}}\over{d\lambda^2}}+\Gamma^{\rho}_{\mu\nu}{{dx^{\mu}}\over
{d\lambda}}{{dx^{\nu}}\over{d\lambda}}+{\cal W}^{-1}g^{\rho\sigma}\partial_y
({\cal W}g_{\sigma\mu}){{dx^{\mu}}\over{d\lambda}}{{dy}\over{d\lambda}}=0,
\label{newgeoeqs1}
\end{equation}
\begin{equation}
{{d^2y}\over{d\lambda^2}}-{1\over 2}\partial_y({\cal W}g_{\mu\nu}){{dx^{\mu}}
\over{d\lambda}}{{dx^{\nu}}\over{d\lambda}}=0.
\label{newgeoeqs2}
\end{equation}
The consistency condition for the geodesic motion in the bulk expressed in 
terms of an affine parameter $\lambda$ [a new parameter $\tilde{\lambda}=
f(\lambda)$] takes the same forms (\ref{metcomp}) [(\ref{georeldtodm1}) and 
(\ref{cstncrel})] except that the metric $g_{\mu\nu}$ now depends on $y$.  

\subsection{Case 1}

In this subsection, we study the geodesic motions of a bulk test particle 
as observed in one lower dimensions with the spacetime metric 
$g_{\mu\nu}(x^{\rho},y)$.  

First, we consider a massless test particle ($\epsilon_D=0$) in the 
bulk spacetime.  For the trivial case of the geodesic motion with 
${{dy}\over{d\lambda}}=0$, the motion is observed in one lower dimensions 
to be that of massless free particle under the influence of gravitational 
field $g_{\mu\nu}$, only.  When ${{dy}\over{d\lambda}}\neq 0$, by using Eqs. 
(\ref{mslsstime}) and (\ref{newgeoeqs2}), one obtains the following relation 
between the two parameters $\lambda$ and $\tilde{\lambda}$:
\begin{equation}
\left({{d\tilde{\lambda}}\over{d\lambda}}\right)^{-1}{d\over{d
\tilde{\lambda}}}\left({{d\tilde{\lambda}}\over{d\lambda}}\right)=
-{1\over 2}{\cal W}^{-{1\over 2}}{\cal W}^{\prime}+{1\over 2}
{\cal W}^{-{1\over 2}}\partial_y({\cal W}g_{\mu\nu}){{dx^{\mu}}\over
{d\tilde{\lambda}}}{{dx^{\nu}}\over{d\tilde{\lambda}}}.
\label{pararel}
\end{equation}
The $x^{\rho}$-component geodesic equation (\ref{newgeoeqs1}) takes 
the following form after Eqs. (\ref{mslsstime}) and (\ref{pararel}) 
are applied:
\begin{equation}
{{d^2x^{\rho}}\over{d\tilde{\lambda}^2}}+\Gamma^{\rho}_{\mu\nu}
{{dx^{\mu}}\over{d\tilde{\lambda}}}{{dx^{\nu}}\over{d\tilde{\lambda}}}=
{1\over 2}{\cal W}^{-{1\over 2}}{\cal W}^{\prime}{{dx^{\rho}}\over
{d\tilde{\lambda}}}-\left[{\cal W}^{-{1\over 2}}g^{\rho\sigma}+{1\over 2}
{\cal W}^{-{1\over 2}}{{dx^{\rho}}\over{d\tilde{\lambda}}}{{dx^{\sigma}}
\over{d\tilde{\lambda}}}\right]{{dx^{\mu}}\over{d\tilde{\lambda}}}
\partial_y({\cal W}g_{\rho\mu}).
\label{newgeoeq}
\end{equation}
The bulk geodesic motion of the massless particle with ${{dy}\over{d\lambda}}
\neq 0$ is therefore observed in one lower dimensions as the motion of a 
massive particle under the additional influence of the extra 
non-gravitational force, unlike the case of the KK zero mode metric 
$g_{\mu\nu}$ as discussed in the previous section.  The extra force 
term on the RHS of Eq. (\ref{newgeoeq}) has both the parallel and the 
perpendicular components given by
\begin{eqnarray}
f^{\rho}_{\parallel}&=&{1\over 2}\left[{\cal W}^{-{1\over 2}}
{\cal W}^{\prime}+{\cal W}^{-{1\over 2}}\partial_y({\cal W}g_{\mu\nu})
{{dx^{\mu}}\over{d\tilde{\lambda}}}{{dx^{\mu}}\over{d\tilde{\lambda}}}
\right]{{dy}\over{d\tilde{\lambda}}}{{dx^{\rho}}\over{d\tilde{\lambda}}},
\cr
f^{\rho}_{\perp}&=&-\left[{\cal W}^{-{1\over 2}}g^{\rho\sigma}+
{\cal W}^{-{1\over 2}}{{dx^{\rho}}\over{d\tilde{\lambda}}}{{dx^{\sigma}}
\over{d\tilde{\lambda}}}\right]{{dx^{\mu}}\over{d\tilde{\lambda}}}
\partial_y({\cal W}g_{\sigma\mu}).
\label{paraperpfrcs}
\end{eqnarray}
These vanish when $g_{\mu\nu}$ is independent of the extra spatial coordinate 
$y$, implying that the extra force is due to the massive KK modes of 
graviton.  

Second, for a massive bulk test particle ($\epsilon_D=1$), by using 
(\ref{affpararel}) one can express the $y$-component bulk geodesic 
equation (\ref{newgeoeqs2}) in terms of $\tilde{\lambda}$ as
\begin{equation}
{{d^2y}\over{d\tilde{\lambda}^2}}-{1\over 2}{{{\cal W}^{\prime}}\over{\cal W}}
\left({{dy}\over{d\tilde{\lambda}}}\right)^2-{1\over 2}\left[{\cal W}-
\left({{dy}\over{d\tilde{\lambda}}}\right)^2\right]{\cal W}^{-1}
\partial_y({\cal W}g_{\mu\nu}){{dx^{\mu}}\over{d\tilde{\lambda}}}
{{dx^{\nu}}\over{d\tilde{\lambda}}}=0.
\label{neq1}
\end{equation}
The $x^{\rho}$-component bulk geodesic equation (\ref{newgeoeqs1}) can be 
expressed in terms of $\tilde{\lambda}$ as follows by applying Eqs. 
(\ref{affpararel}) and (\ref{neq1}):
\begin{equation}
{{d^2x^{\rho}}\over{d\tilde{\lambda}^2}}+\Gamma^{\rho}_{\mu\nu}
{{dx^{\mu}}\over{d\tilde{\lambda}}}{{dx^{\nu}}\over{d\tilde{\lambda}}}=
{1\over 2}{{{\cal W}^{\prime}}\over{\cal W}}{{dy}\over{d\tilde{\lambda}}}
{{dx^{\rho}}\over{d\tilde{\lambda}}}-
{\cal W}^{-1}\left[g^{\rho\sigma}+{1\over 2}{{dx^{\rho}}\over
{d\tilde{\lambda}}}{{dx^{\sigma}}\over{d\tilde{\lambda}}}\right]
{{dy}\over{d\tilde{\lambda}}}{{dx^{\mu}}\over{d\tilde{\lambda}}}
\partial_y\left({\cal W}g_{\sigma\mu}\right).
\label{neq2}
\end{equation}
So, the bulk geodesic motion of a massive test particle is observed in 
one lower dimensions as the motion of a massive particle under the additional 
influence of the extra non-gravitational force, also unlike the case 
of $y$-independent $g_{\mu\nu}$.  The extra force term on the 
RHS of Eq. (\ref{neq2}) has both the parallel and the perpendicular 
components given by
\begin{eqnarray}
f^{\rho}_{\parallel}&=&{1\over 2}\left[{{{\cal W}^{\prime}}\over{\cal W}}
+{\cal W}^{-1}\partial_y({\cal W}g_{\mu\nu}){{dx^{\mu}}\over{d\tilde{\lambda}}}
{{dx^{\mu}}\over{d\tilde{\lambda}}}\right]{{dy}\over{d\tilde{\lambda}}}
{{dx^{\rho}}\over{d\tilde{\lambda}}},
\cr
f^{\rho}_{\perp}&=&-{\cal W}^{-1}\left[g^{\rho\sigma}+{{dx^{\rho}}\over
{d\tilde{\lambda}}}{{dx^{\sigma}}\over{d\tilde{\lambda}}}\right]
{{dy}\over{d\tilde{\lambda}}}{{dx^{\mu}}\over{d\tilde{\lambda}}}
\partial_y({\cal W}g_{\sigma\mu}).
\label{frceoncst}
\end{eqnarray}
As expected, these vanish when $g_{\mu\nu}$ is independent of $y$.  

We now obtain the expression for the extra $(D-1)$-force $F^{\mu}$ acting 
on the particle from the $(D-1)$-dimensional perspective.  The bulk test 
particle of mass $m_0$ is observed to have mass $\tilde{m}_0$ given 
by Eq. (\ref{ffmss}) from the perspective of $(D-1)$-dimensional spacetime 
with the metric $g_{\mu\nu}$.  By using Eq. (\ref{neq1}), one obtains the 
following rate of mass change with $\tilde{\lambda}$ as observed in 
the embedded $(D-1)$-dimensional spacetime:
\begin{equation}
{{d\tilde{m}_0}\over{d\tilde{\lambda}}}=-{{\tilde{m}_0}\over{2}}\left[
{\cal W}^{-1}{\cal W}^{\prime}-{\cal W}^{-1}\partial_y({\cal W}g_{\mu\nu})
{{dx^{\mu}}\over{d\tilde{\lambda}}}{{dx^{\nu}}\over{d\tilde{\lambda}}}\right]
{{dy}\over{d\tilde{\lambda}}}.
\label{mssch3}
\end{equation}
From Eqs. (\ref{neq2}) and (\ref{mssch3}), we obtain the following equation 
describing the conservation of energy and the Newton's second law of 
mechanics in $(D-1)$-dimensions:
\begin{equation}
{{Dp^{\mu}}\over{d\tilde{\lambda}}}\equiv {{dp^{\mu}}\over{d\tilde{\lambda}}}
+\Gamma^{\mu}_{\rho\sigma}{{dx^{\rho}}\over{d\tilde{\lambda}}}p^{\sigma}=
-\tilde{m}_0{\cal W}^{-1}g^{\mu\nu}{{dx^{\rho}}\over{d\tilde{\lambda}}}
\partial_y({\cal W}g_{\nu\rho}){{dy}\over{d\tilde{\lambda}}}.
\label{ceflw3}
\end{equation}
The extra force $F^{\mu}$ on the RHS of Eq. (\ref{ceflw3}) has both the 
parallel and orthogonal components given by
\begin{eqnarray}
F^{\mu}_{\parallel}&=&\tilde{m}_0{\cal W}^{-1}{{dx^{\rho}}\over
{d\tilde{\lambda}}}{{dx^{\sigma}}\over{d\tilde{\lambda}}}\partial_y
({\cal W}g_{\rho\sigma}){{dy}\over{d\tilde{\lambda}}}{{dx^{\mu}}\over
{d\tilde{\lambda}}},
\cr
F^{\mu}_{\perp}&=&-\tilde{m}_0{\cal W}^{-1}\left[g^{\mu\nu}+
{{dx^{\mu}}\over{d\tilde{\lambda}}}{{dx^{\nu}}\over{d\tilde{\lambda}}}\right]
{{dy}\over{d\tilde{\lambda}}}{{dx^{\rho}}\over{d\tilde{\lambda}}}
\partial_y({\cal W}g_{\nu\rho}).
\label{extrf3}
\end{eqnarray}
As expected, when $g_{\mu\nu}$ does not depend on $y$, $F^{\mu}_{\perp}$ 
vanishes and $F^{\mu}_{\parallel}$ takes the form of the RHS of Eq. 
(\ref{frcl}).  

Even with the choice of the metric by $g_{\mu\nu}$ for the $(D-1)$-dimensional 
spacetime, which it has been previously regarded as the natural canonical 
choice for the $(D-1)$-dimensional spacetime embedded in brane worlds, the 
motion of the particle observed in the $(D-1)$-dimensional spacetime appears 
to be under the additional influence of the abnormal force term, which cannot 
be explained by laws of physics in $(D-1)$-dimensions, if the metric 
$g_{\mu\nu}$ depends on the extra spatial coordinate $y$.  Namely, the 
equations (\ref{newgeoeq}) and (\ref{neq2}) describing the particle 
trajectory $x^{\mu}(\tilde{\lambda})$ observed in one lower dimensions 
have nonzero parallel component force term $f^{\rho}_{\parallel}$ and 
the the mass change (\ref{mssch3}) with $\tilde{\lambda}$ is not in 
accordance with the conventional formula (\ref{relwthcht}) for a given 
extra non-gravitational force $F^{\mu}=-\tilde{m}_0{\cal W}^{-1}g^{\mu\nu}
{{dx^{\rho}}\over{d\tilde{\lambda}}}\partial_y({\cal W}g_{\nu\rho}){{dy}
\over{d\tilde{\lambda}}}$.  Also, with a choice of the metric frame 
$\tilde{g}_{\mu\nu}$, the same holds true as expected, as we will see in 
the following subsection.  Note, the abnormal force term is not due to the 
wrong choice
\footnote{If one chooses a non-affine parameter to describe the motion of a 
particle, the abnormal force term also occurs in the equation for particle 
trajectory.  To see this, we consider the following geodesic 
equation for a free particle, whose motion is under the influence of the 
gravitational force, only:
\begin{equation}
{{d^2x^{\rho}}\over{ds^2}}+\Gamma^{\rho}_{\mu\nu}{{dx^{\mu}}\over{ds}}
{{dx^{\nu}}\over{ds}}=0,
\label{geoeqs}
\end{equation}
where $s$ is an affine parameter.  If we take a new parameter $\tilde{s}=
f(s)$ to parameterize the motion of the particle, then the above geodesic 
equations transform to the following form:
\begin{equation}
{{d^2x^{\rho}}\over{d\tilde{s}^2}}+\Gamma^{\rho}_{\mu\nu}{{dx^{\mu}}\over
{d\tilde{s}}}{{dx^{\nu}}\over{d\tilde{s}}}=-{{d^2\tilde{s}/ds^2}\over
{(d\tilde{s}/ds)^2}}{{dx^{\rho}}\over{d\tilde{s}}}.
\label{ngeoeqs}
\end{equation}
So, through non-affine transformation one induces an extra velocity dependent 
fictitious force term $-{{d^2\tilde{s}/ds^2}\over{(d\tilde{s}/ds)^2}}
{{dx^{\rho}}\over{d\tilde{s}}}$ parallel to the four-velocity ${{dx^{\rho}}
\over{d\tilde{s}}}$ of the particle.  Eq. (\ref{ngeoeqs}) also shows that 
the geodesic equations (\ref{geoeqs}) are invariant only under the affine 
transformations $s\to\tilde{s}=as+b$, $a,b\in{\bf R}$.} 
of parameter $\tilde{\lambda}$ describing motion observed in 
$(D-1)$-dimensions, since we have fixed (up to affine transformations) the 
parameter $\tilde{\lambda}$ through the $(D-1)$-dimensional affine 
conditions (\ref{naffin1}) and (\ref{naffin2}).  
So, the massive KK modes of graviton not only give a small correction 
to Newton's $1/r^2$ law of four-dimensional gravity but also causes violation 
of four-dimensional laws of physics, which can be an indication that our 
four-dimensional world is embedded in higher-dimensional spacetime.  

If we take the viewpoint that laws of physics in $(D-1)$-dimensional 
spacetime should not be violated, then we have to take the metric 
$\bar{g}_{\mu\nu}$ for which the equation for the particle trajectory 
does not have abnormal force term as the physical metric of the 
$(D-1)$-dimensional spacetime.  With a choice of such metric, according to 
Eq. (\ref{relwthcht}), the $(D-1)$-dimensional mass $\bar{m}_0$ should 
change with an affine parameter $\bar{\lambda}$ as
\begin{equation}
{{d\bar{m}_0}\over{d\bar{\lambda}}}=-\bar{g}_{\mu\nu}F^{\mu}{{dx^{\nu}}\over
{d\bar{\lambda}}},
\label{trmassch}
\end{equation}
where $\bar{m}_0=m_0{{d\bar{\lambda}}\over{d\lambda}}$, the parameter 
$\bar{\lambda}$ is defined through the relation $\bar{g}_{\mu\nu}
{{dx^{\mu}}\over{d\bar{\lambda}}}{{dx^{\nu}}\over{d\bar{\lambda}}}=-1$,  
$F^{\mu}$ is the extra non-gravitational force observed in $(D-1)$-dimensions, 
and of course the parameter $\lambda$ is defined through Eq. 
(\ref{metcomp}) with $\epsilon_D=1$.  
In the case where $g_{\mu\nu}$ is independent of $y$, such physical metric 
is given by $\bar{g}_{\mu\nu}=g_{\mu\nu}$.  For a general $y$-dependent 
$g_{\mu\nu}$, it seems not clear whether a simple and natural form of the 
physical metric $\bar{g}_{\mu\nu}$ that satisfies this equation exists.  

\subsection{Case 2}

In this subsection, we study the geodesic motion of a bulk test particle 
as observed in the (comoving) hypersurface $y=y(\tilde{\lambda})$ with the 
metric $\tilde{g}_{\mu\nu}={\cal W}(y)g_{\mu\nu}(x^{\rho},y)$.  

First, we consider the case of a free massless bulk particle ($\epsilon_D=0$).
When ${{dy}\over{d\lambda}}=0$, one can put the $x^{\rho}$-component 
geodesic equation (\ref{newgeoeqs1}) to the form (\ref{nultonul2}).  
When ${{dy}\over{d\lambda}}\neq 0$, by using Eqs. (\ref{nulltotime}) and 
(\ref{newgeoeqs2}), one can see that the affine parameters $\lambda$ and 
$\tilde{\lambda}$ are related as
\begin{equation}
\left({{d\tilde{\lambda}}\over{d\lambda}}\right)^{-1}{d\over{d\tilde{\lambda}}}
\left({{d\tilde{\lambda}}\over{d\lambda}}\right)={1\over 2}\partial_y
\left({\cal W}g_{\mu\nu}\right){{dx^{\mu}}\over{d\tilde{\lambda}}}
{{dx^{\nu}}\over{d\tilde{\lambda}}}.
\label{lambtillamb}
\end{equation}
Applying Eqs. (\ref{nulltotime}) and (\ref{lambtillamb}), one can put the 
$x^{\rho}$-component bulk geodesic equation (\ref{newgeoeqs1}) into the 
following form in terms of the new parameter $\tilde{\lambda}$:
\begin{equation}
{{d^2x^{\rho}}\over{d\tilde{\lambda}^2}}+\tilde{\Gamma}^{\rho}_{\mu\nu}
{{dx^{\mu}}\over{d\tilde{\lambda}}}{{dx^{\nu}}\over{d\tilde{\lambda}}}=
-\left[{\cal W}^{-1}g^{\rho\sigma}+{1\over 2}{{dx^{\rho}}\over
{d\tilde{\lambda}}}{{dx^{\sigma}}\over{d\tilde{\lambda}}}\right]
{{dx^{\mu}}\over{d\tilde{\lambda}}}\partial_y\left({\cal W}g_{\sigma\mu}
\right),
\label{xrhoeqn}
\end{equation}
where $\tilde{\Gamma}^{\rho}_{\mu\nu}$ is the Christoffel symbol for the 
metric $\tilde{g}_{\mu\nu}={\cal W}(y)g_{\mu\nu}(x^{\rho},y)$ and we used 
the fact that $\tilde{\Gamma}^{\rho}_{\mu\nu}=\Gamma^{\rho}_{\mu\nu}$.  So, 
the geodesic motion of a massless free particle in the bulk spacetime with the 
metric (\ref{bulkmet}) is observed on the (comoving) hypersurface $y=
y(\tilde{\lambda})$ as the motion of a massive particle under the additional 
influence of the extra non-gravitational force.
Unlike the case of the $y$-independent $g_{\mu\nu}$, the extra force term on 
the RHS of Eq. (\ref{xrhoeqn}) is no longer parallel to the $(D-1)$-velocity 
${{dx^{\rho}}\over{d\tilde{\lambda}}}$ of the test particle.  The parallel 
and the perpendicular components of the extra force term are given by
\begin{eqnarray}
f^{\rho}_{\parallel}&=&{1\over 2}\partial_y\left({\cal W}g_{\mu\nu}\right)
{{dx^{\mu}}\over{d\tilde{\lambda}}}{{dx^{\nu}}\over{d\tilde{\lambda}}}
{{dx^{\rho}}\over{d\tilde{\lambda}}},
\cr
f^{\rho}_{\perp}&=&-\left[{{dx^{\rho}}\over{d\tilde{\lambda}}}
{{dx^{\sigma}}\over{d\tilde{\lambda}}}+{\cal W}^{-1}g^{\rho\sigma}
\right]{{dx^{\mu}}\over{d\tilde{\lambda}}}\partial_y\left({\cal W}g_{\sigma\mu}
\right).
\label{decompfrc1}
\end{eqnarray}
As expected, the perpendicular component $f^{\rho}_{\perp}$ vanishes and 
the parallel component $f^{\rho}_{\parallel}$ takes the form of the RHS of 
Eq. (\ref{xrhoeqnull}), when $g_{\mu\nu}$ is independent of $y$, implying 
that the perpendicular component is due to the massive KK modes of graviton.  

Second, we consider the case of a massive bulk test particle 
($\epsilon_D=1$).  By using Eq. (\ref{affnpararel2}), one can put the 
$y$-component geodesic equations (\ref{newgeoeqs2}) into the following form 
in terms of a new parameter $\tilde{\lambda}$:
\begin{equation}
{{d^2y}\over{d\tilde{\lambda}^2}}-{1\over 2}\left[1-\left({{dy}\over
{d\tilde{\lambda}}}\right)^2\right]\partial_y\left({\cal W}g_{\mu\nu}
\right){{dx^{\mu}}\over{d\tilde{\lambda}}}{{dx^{\nu}}\over
{d\tilde{\lambda}}}=0.
\label{nngeoeqs2}
\end{equation} 
The $x^{\rho}$-component bulk geodesic equation (\ref{newgeoeqs1}) takes the 
following form after Eqs. (\ref{affnpararel2}) and (\ref{nngeoeqs2}) are 
applied:
\begin{equation}
{{d^2x^{\rho}}\over{d\tilde{\lambda}^2}}+\tilde{\Gamma}^{\rho}_{\mu\nu}
{{dx^{\mu}}\over{d\tilde{\lambda}}}{{dx^{\nu}}\over{d\tilde{\lambda}}}=
-\left[{\cal W}^{-1}g^{\rho\sigma}+{1\over 2}{{dx^{\rho}}\over
{d\tilde{\lambda}}}{{dx^{\sigma}}\over{d\tilde{\lambda}}}\right]
{{dy}\over{d\tilde{\lambda}}}{{dx^{\mu}}\over{d\tilde{\lambda}}}
\partial_y\left({\cal W}g_{\sigma\mu}\right).
\label{nngeoeqs1}
\end{equation}
So, the geodesic motion of a massive particle in the bulk spacetime with 
the metric (\ref{bulkmet}) is observed on the (comoving) hypersurface 
$y=y(\tilde{\lambda})$ as the motion of a massive particle under the 
additional influence of the extra non-gravitational force.   
As in the massless test particle case, the extra force term on the RHS of 
Eq. (\ref{nngeoeqs1}) has both parallel and perpendicular components given by
\begin{eqnarray}
f^{\rho}_{\parallel}&=&{1\over 2}\partial_y\left({\cal W}g_{\mu\nu}\right)
{{dy}\over{d\tilde{\lambda}}}{{dx^{\mu}}\over{d\tilde{\lambda}}}
{{dx^{\nu}}\over{d\tilde{\lambda}}}{{dx^{\rho}}\over{d\tilde{\lambda}}},
\cr
f^{\rho}_{\perp}&=&-\left[{{dx^{\rho}}\over{d\tilde{\lambda}}}
{{dx^{\sigma}}\over{d\tilde{\lambda}}}+{\cal W}^{-1}g^{\rho\sigma}
\right]{{dy}\over{d\tilde{\lambda}}}
{{dx^{\mu}}\over{d\tilde{\lambda}}}\partial_y\left({\cal W}g_{\sigma\mu}
\right).
\label{decompfrc2}
\end{eqnarray}
As expected, when $g_{\mu\nu}$ is independent of $y$, $f^{\rho}_{\parallel}$ 
takes the form of the RHS of Eq. (\ref{xdireceqs}) and $f^{\rho}_{\perp}$ 
vanishes.  

We now obtain the expression for the extra non-gravitational $(D-1)$-force 
$F^{\mu}$ observed on the (comoving) hypersurface $y=y(\tilde{\lambda})$.  
A bulk test particle with mass $m_0$ is measured on the hypersurface 
$y=y(\tilde{\lambda})$ to have mass $\tilde{m}_0$ given by Eq. 
(\ref{ffmass2}).  By using Eq. (\ref{nngeoeqs2}), we obtain the following 
mass change with $\tilde{\lambda}$ as observed on the hypersurface:  
\begin{equation}
{{d\tilde{m}_0}\over{d\tilde{\lambda}}}={{\tilde{m}_0}\over{2}}\partial_y
({\cal W}g_{\mu\nu}){{dx^{\mu}}\over{d\tilde{\lambda}}}
{{dx^{\nu}}\over{d\tilde{\lambda}}}{{dy}\over{d\tilde{\lambda}}}.
\label{mssch4}
\end{equation}
So, from Eqs. (\ref{nngeoeqs1}) and (\ref{mssch4}), we obtain the equation 
${{Dp^{\mu}}\over{d\tilde{\lambda}}}=F^{\mu}$ which has the same form 
(\ref{ceflw3}) as the case 1.  However, since we have chosen different 
$(D-1)$-dimensional metric, the expressions for the parallel and the 
perpendicular components of $F^{\mu}$ are instead given by
\begin{eqnarray}
F^{\mu}_{\parallel}&=&\tilde{m}_0{{dx^{\rho}}\over{d\tilde{\lambda}}}
{{dx^{\sigma}}\over{d\tilde{\lambda}}}\partial_y({\cal W}g_{\rho\sigma})
{{dy}\over{d\tilde{\lambda}}}{{dx^{\mu}}\over{d\tilde{\lambda}}},
\cr
F^{\mu}_{\perp}&=&-\tilde{m}_0\left[{\cal W}^{-1}g^{\mu\nu}+
{{dx^{\mu}}\over{d\tilde{\lambda}}}{{dx^{\nu}}\over{d\tilde{\lambda}}}\right]
{{dy}\over{d\tilde{\lambda}}}{{dx^{\rho}}\over{d\tilde{\lambda}}}
\partial_y({\cal W}g_{\nu\rho}).
\label{fccomp4}
\end{eqnarray}
As expected, when $g_{\mu\nu}$ is independent of $y$, $F^{\mu}_{\perp}$ 
vanishes and $F^{\mu}_{\parallel}$ takes the form of the RHS of Eq. 
(\ref{phyleq}).

\section{Conclusions}

In this paper, we carefully studied the geodesic motions of a test particle 
in the bulk spacetime of general gravitating configurations in the RS 
scenario as observed in the embedded spacetime of one lower dimensions. 
We presented the explicit equations describing such particle motion 
perceived by an observer in one lower dimensions and the explicit forms 
of the extra force on the particle measured in one lower dimensions.  
Such equations and extra forces are inconsistent with laws of particle 
mechanics in one lower dimensions.  Such inconsistency does not mean the 
violation of physics in one lower dimensions, but results from our effort 
to interpret the physical process which is higher-dimensional in nature 
with physics of one lower dimensions.  The RS model assumes that the extra 
spatial dimension is noncompact, and therefore generically physical phenomena 
in the RS model have to show higher-dimensional character, which is 
observed to be inconsistent with physics of our four-dimensional world.  
So, one can test the RS scenario by detecting inconsistency with the 
four-dimensional physics such as the one present in this paper.  However, 
one has to note that since the current RS scenario assumes that the 
motion of ordinary matter in our (visible) universe is confined within the 
TeV brane, which is assumed to be at {\it fixed} distance from the Planck 
brane, due to yet unknown non-gravitational mechanism, the extra force 
discussed in this paper will not be detected by the lower-dimensional 
observers on the (visible) TeV brane of the current RS model.  
The extra force discussed in this paper can be measured only by an observer 
who follows a test particle with the nonzero velocity component along the 
extra spatial direction.

\end{document}